\documentclass{PoS}

\usepackage{enumerate}
\usepackage{amsmath,amssymb}
\usepackage{mathrsfs}
\usepackage{bm}

\usepackage{ulem}
\usepackage{cleveref}

\title{
\begin{picture}(0,0)(0,0)%
 \put(360,105){\makebox(0,0)[l]{\textnormal{\normalsize
 CHIBA-EP-243}}}%
\end{picture}%
Complex poles, spectral function and reflection positivity violation of Yang-Mills theory 
} 

\ShortTitle{Complex poles, spectral function and reflection positivity violation}

\author{\speaker{Kei-Ichi Kondo}\\
Department of Physics, Graduate School of Science, Chiba University, Chiba 263-8522, Japan
\\
E-mail: \email{kondok@faculty.chiba-u.jp }
}

\author{Yui Hayashi\\
Department of Physics, 
Graduate School of Science and Engineering, 
Chiba University, Chiba 263-8522, Japan
}

\author{Ryutaro Matsudo\\
Department of Physics, 
Graduate School of Science and Engineering, 
Chiba University, Chiba 263-8522, Japan
}

\author{Yutaro Suda\\
Department of Physics, 
Graduate School of Science, 
Chiba University, Chiba 263-8522, Japan
}

\author{Masaki Watanabe\\
Department of Physics, 
Graduate School of Science and Engineering, 
Chiba University, Chiba 263-8522, Japan
}

\abstract{
We discuss the analytic continuation of the gluon propagator from the Euclidean region to the complex squared-momentum plane towards the Minkowski region from a viewpoint of gluon confinement. 
For this purpose, we investigate the massive Yang-Mills model with one-loop quantum corrections, which is to be identified with a low-energy effective theory of the Yang-Mills theory in the sense that the confining decoupling solution for the Euclidean gluon and ghost propagators of the Yang-Mills theory in the Landau gauge obtained by numerical simulations on the lattice are reproduced with good accuracy from the massive Yang-Mills model by taking into account one-loop quantum corrections. 
We show that the gluon propagator in the massive Yang-Mills model has a pair of complex conjugate poles or ``tachyonic''  poles of multiplicity two, in accordance with the fact that the gluon field has a negative spectral function, while the ghost propagator has at most one ``unphysical'' pole. 
These results are consistent with  general relationships between the number of complex poles of a propagator and the sign of the spectral function originating from the branch cut in the Minkowski region under some assumptions on the asymptotic behaviors of the propagator. 
Consequently, we give an analytical proof for violation of the reflection positivity as a necessary condition for gluon confinement for any choice of the parameters in the massive Yang-Mills model, including the physical point. 
Moreover, the complex structure of the propagator enables us to explain why the gluon propagator in the Euclidean region is well described by the Gribov-Stingl form.

}

\FullConference{Light Cone 2019 - QCD on the light cone: from hadrons to heavy ions - LC2019\\
		16-20 September 2019\\
		Ecole Polytechnique, Palaiseau, France}

\begin{document}

\section{.Introduction.}


\begin{figure}
\centering
\includegraphics[width=7.5cm]{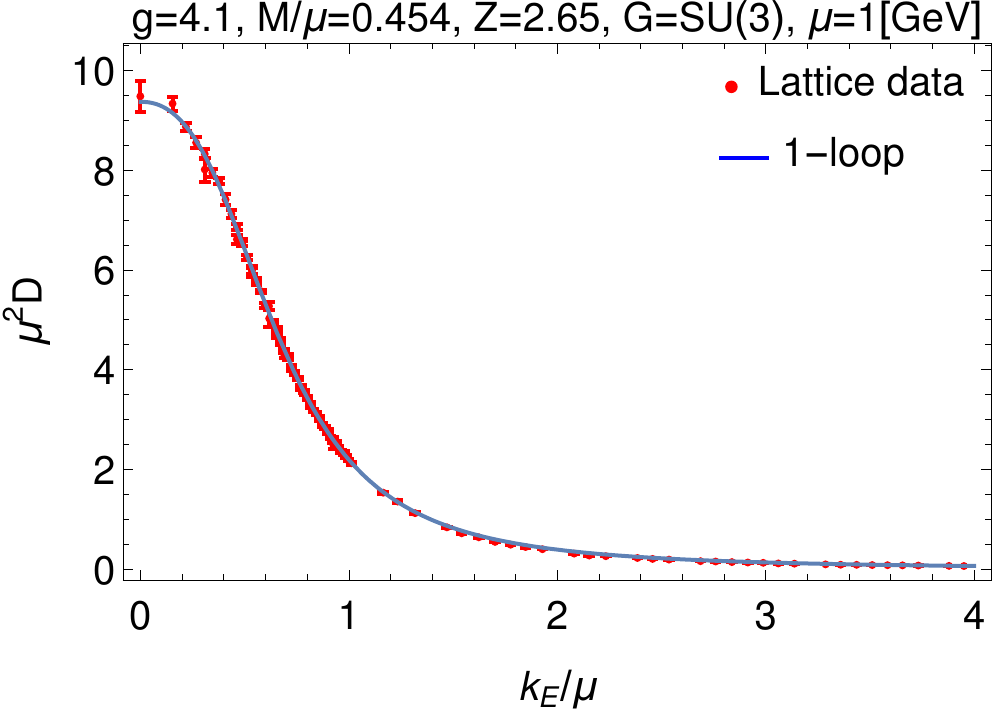}
\includegraphics[width=7.5cm]{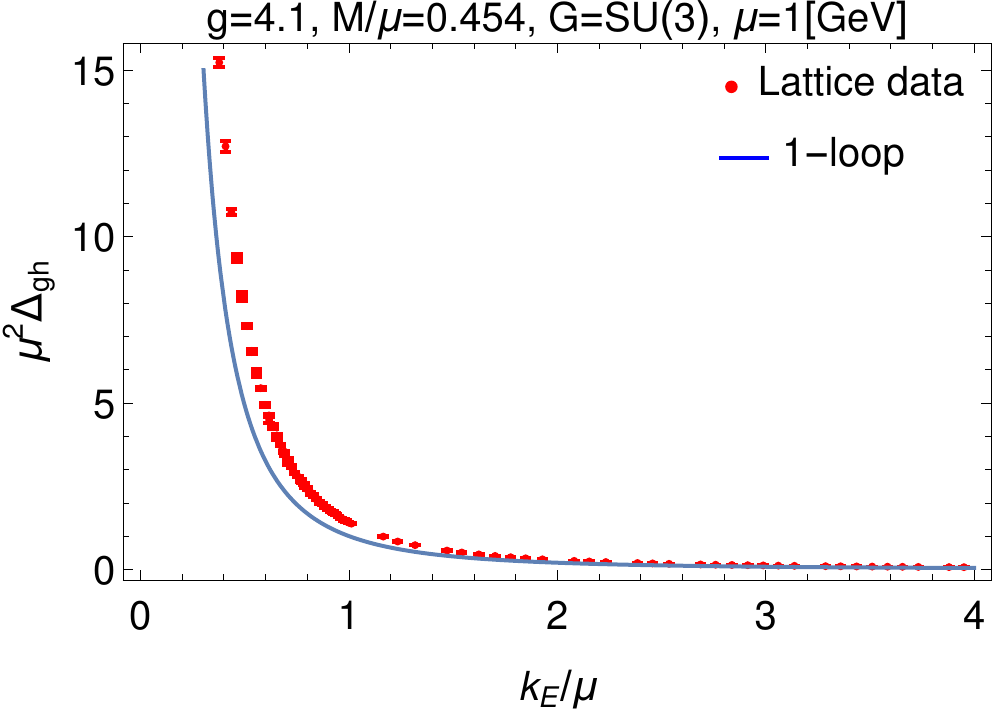}
\vskip -0.5cm
\caption{\small
The gluon propagator $\mathscr{D}$ and ghost propagator $\Delta_{gh}$ as functions of the Euclidean momentum $k_E$.
}
\label{figFitGluonRaw}
\end{figure}

In this talk we consider the $D$-dimensional Yang-Mills theory in the manifestly Lorentz covariant gauge of the 
Lorenz type  from the viewpoint of confinement. 
The Euclidean gluon propagator and ghost propagator in the Landau gauge $\partial_\mu \mathscr{A}_\mu=0$ are written as 
\begin{equation}
   \mathscr{D}_{AB}^{\mu\nu} (k_E) = \delta^{AB} \left( \delta^{\mu\nu} - \frac{k_E^\mu k_E^\nu}{k_E^2}\right) \mathscr{D}(k_E^2) , \quad 
   \Delta^{gh}_{AB} (k_E) = \delta^{AB}  \Delta_{gh}(k_E^2) .
\end{equation}
We focus on the small momentum region. 
In the end of the last century, 
the \textbf{scaling solution} which is consistent with the Gribov/Zwanziger prediction  [realized only for $D=2$] was found:
\begin{align}
 \mathscr{D}(k_E^2) \sim (k_E^2)^{\alpha-1} \downarrow 0 , \quad \Delta_{gh}(k_E^2) \sim \frac{Z_{gh}}{(k_E^2)^{1-\beta}} \uparrow \infty \quad (1<\alpha=-2\beta<2) \quad \text{as} \quad  k_E^2 \downarrow 0  .
\end{align}
After around 2006, the  \textbf{decoupling solution} which exhibits massive gluon and massless ghost was recognized to be a true confining solution for $D=4,3$ (See Figure~\ref{figFitGluonRaw})
\begin{align}
 \mathscr{D}(k_E^2) := \frac{F(k_E^2)}{k_E^2} \to {\rm const. }, \quad \Delta_{gh}(k_E^2) := \frac{G(k_E^2)}{k_E^2} \sim \frac{Z_{gh}}{k_E^2} \uparrow \infty \quad \text{as} \quad  k_E^2 \downarrow 0  .
\end{align}


In order to understand these solutions in the Yang-Mills theory, we consider the \textbf{massive Yang-Mills model} of Curci-Ferrari type 
 described by the ordinary  massless Yang-Mills (YM) Lagrangian in the  manifestly Lorentz covariant gauge of the Lorenz  type with the  gauge-fixing (GF) term and the associated Faddeev-Popov (FP) ghost term plus a   naive gluon mass term,
\begin{align}
& \mathscr{L}_{\rm mYM} 
=   
\mathscr{L}_{\rm YM} + \mathscr{L}_{\rm GF} + \mathscr{L}_{\rm FP} + \mathscr{L}_{\rm m}, \quad
\nonumber\\
& \mathscr{L}_{\rm YM} = -\frac{1}{4} \mathscr{F}^{\mu\nu A}   \mathscr{F}_{\mu\nu}^A ,  
\quad   \mathscr{L}_{\rm GF} = \mathscr{N}^A  \partial^\mu \mathscr{A}_{\mu}^A + \frac{\alpha}{2}\mathscr{N}^A\mathscr{N}^A 
\to -\frac{1}{2}\alpha^{-1} (\partial^\mu \mathscr{A}^A_\mu)^2 
\nonumber\\
& \mathscr{L}_{\rm FP} = i\mathscr{\bar C}^A \partial^\mu \mathscr{D}_\mu[\mathscr{A}]^{AB} \mathscr{C}^B 
= i\mathscr{\bar{C}}^A\partial^\mu(\partial_\mu\mathscr{C}^A+gf_{ABC}\mathscr{A}^B_\mu\mathscr{C}^C) ,
\quad \mathscr{L}_{\rm m} = \frac12 M^2 \mathscr{A}^{\mu A}   \mathscr{A}_\mu^A .
\end{align}
Here  $g, M$ and $\alpha (\to 0)$ are the parameters of the massive Yan-Mills model.
\noindent
In this talk we regard the massive Yang-Mills model as a low-energy effective theory for describing the $D=4$ \textbf{decoupling solution of the Yang-Mills theory} to examine \textbf{gluon and quark confinement}. 
In the Euclidean region, the \textbf{massive Yang-Mills model with (at least  one-loop) quantum corrections} being included well reproduces propagators and vertices of the decoupling solution in the covariant Landau gauge in the confining phase of the Yang-Mills theory,  as initiated by Tissier and Wschebor  \cite{TW10} and  demonstrated in the last ten years.  
This can be done in the so-called \textbf{infrared safe renormalization scheme}  by taking the renormalization conditions resulting from the \textbf{non-renormalization theorem}. 
Both gluon propagator and ghost propagator in the decoupling solution of the Yang-Mills theory in the Landau gauge are well reproduced by the massive Yang-Mills model at the specific values of parameters $g$ and $M$, which we call the \textbf{physical point} for the Yang-Mills theory \cite{KWHMS19}:
\begin{align}
  g = 4.1 \pm 0.1 , \ \ 
  \frac{M}{\mu} = 0.454 \pm 0.004 . \ 
\label{exFitParams-s}
\end{align}
%
See Fig.~\ref{figFitGluonRaw}.
For the decoupling solution, the running coupling constant is always finite and asymptotic free in the infrared as well as the ultraviolet. 
The origin of such a gluon mass term can be discussed separately. 




\section{Reflection positivity violation in the Euclidean region
}
\setcounter{equation}{0}
\normalsize 

Reflection positivity is one of the 
the \textbf{Osterwalder-Schrader (OS) axioms}, 
 general properties to be satisfied for the \textbf{Euclidean quantum field theory} formulated in the Euclidean space: 
\\
(OS.3) {\bf Reflection positivity }:
Any complex-valued test function $f_0 \in \mathbb{C}_1,f_1 \in \mathscr{S}_{+}(\mathbb{R}^{D}),\cdots,f_N \in \mathscr{S}_{+}(\mathbb{R}^{DN})$, the Euclidean Green functions  $S_{n+m} $ satisfy
%
\begin{eqnarray}
\sum_{n,m = 0}^{N} S_{n+m} (x_1,\cdots,x_n,x_{n+1},\cdots,x_{n+m})
 f_n(\theta x_n,\theta x_{n-1},\cdots,\theta x_1)^*
f_m( x_{n+1},\cdots, x_{n+m}) \geq 0 ,
\nonumber
\end{eqnarray}
where 
$\mathscr{S}_{+}(\mathbb{R}^D)$ denotes  a complex-valued test (Schwartz) function with support in $\{ (\bm{x},x_D); x_D >0\}$ 
and $\theta$ is a \textbf{reflection} with respect to a hyperplane  $x^0 = 0$: for a function $f_n \in \mathscr{S}(\mathbb{R}^{Dn})$, 
\begin{equation}
\theta x = \theta (x^0,\boldsymbol{x}) =   (-x^0,\boldsymbol{x}) , \ 
(\theta f_n)(x_1,\cdots,x_n) = f_n(\theta x_1,\cdots,\theta x_n) .
\end{equation}
This is a Euclidean version of the positivity axiom in the \textbf{Wightman axioms}  for the \textbf{relativistic quantum field theory} formulated in the Minkowski spacetime.

(W.3) {\bf Positivity}: For all $f_0 \in \mathbb{C}_1,f_1 \in \mathscr{S}(\mathbb{R}^{D}),\cdots,f_N \in \mathscr{S}(\mathbb{R}^{DN})$,
$(N = 0,1,2,\cdots)$
\begin{align}
& \sum_{n,m = 0}^{N} W_{n+m} (x_1,\cdots,x_n,x_{n+1},\cdots,x_{n+m})
  f_n( x_n, x_{n-1},\cdots,x_1)^{*} f_m( x_{n+1},\cdots, x_{n+m})  \geq 0 .
\nonumber
\end{align}
The \textbf{violation of reflection positivity} in the Euclidean region is regarded as a \textbf{necessary condition for gluon confinement}. 
To demonstrate the \textbf{violation of reflection positivity}, one counterexample suffices.
We focus on a special case ($N=2$) of a single propagator $S_2=\mathscr{D}$. Then the reflection positivity reads
\begin{align}
  & \int d^Dx \int d^Dy f^*(\bm{x},-x_D) \mathscr{D}(\bm{x}-\bm{y},x_D-y_D) f(\bm{y},y_D) \ge 0 , \
 f \in \mathscr{S}_{+}(\mathbb{R}^D) ,
\end{align}
which is rewritten as 
\begin{align}
 \int_{0}^{\infty} dt \int_{0}^{\infty} dt^\prime \int d^{D-1}\bm{p} f^*(\bm{p},t) f(\bm{p},t^\prime) \Delta(\bm{p}, -(t+t^\prime))
 \ge 0,
 \label{RP1}
\end{align}
where we defined the \textbf{Schwinger function} $\Delta(\bm{p},x_D-y_D)$ by
\begin{align}
\mathscr{D}(x-y) := \int d^{D-1}\bm{p} \ e^{i \bm{p} \cdot (\bm{x}-\bm{y})} \Delta(\bm{p},x_D-y_D)  
 .
\end{align}
For the inequality (\ref{RP1}) to hold for any  $f \in \mathscr{S}_{+}(\mathbb{R}^D)$, the Schwinger function $\Delta$ must satisfy 
\begin{align}
 \Delta(\bm{p}, -(t+t^\prime)) = \Delta(\bm{p},  t+t^\prime )  \ge 0   
 .
\end{align}
We consider a specific \textbf{Schwinger function} defined by the Fourier transform of the propagator:
\begin{equation}
\Delta(t) := \Delta(\bm{p},t)|_{\bm{p}=0}
:= \int d^{D-1}x   e^{-i\bm{p} \cdot \bm{x}}  \mathscr{D}(\bm{x},t)|_{\bm{p}=0}
= \int_{-\infty}^{+\infty} \frac{dp_{E}^D}{2\pi}  e^{ip_{E}^D t} \tilde{\mathscr{D}} (\bm{p}=\bm{0} ,p_{E}^D)  .
\end{equation} 

For the  free massive theory ($g=0$), we find $\Delta(t)$ is positive for any $t$: 
\begin{align}
  \tilde{\mathscr{D}}(p) = \frac{1}{p^2+m^2}  \quad (m>0)
\Longrightarrow
  \Delta(t) 
=  \int_{-\infty}^{+\infty} \frac{dp_D}{2\pi}  e^{ip_D t} \frac{1}{p_D^2+m^2} 
= \frac{1}{2m} e^{-m|t|} > 0  
 .
\end{align}
There is  no reflection-positivity violation for the free massive propagator, as expected. 
For unconfined particles, the reflection positivity should hold. 

\textit{The reflection positivity is violated for the massive Yang-Mills model at the physical point of parameters}, as already shown by the numerical calculations.
We proceed to prove that the reflection positivity is violated for any choice of the parameters in the massive Yang-Mills model. 
This suggests the \textit{reflection positivity violation for the decoupling solution of the Yang-Mills theory}.  
In order to consider the origin, we proceed the complex analysis of the Yang-Mills theory.

\section{Complex analysis of the gluon propagator}

In the Minkowski region with time-like momentum $k^2>0$, a propagator $\mathscr{D}(k^2)$  has the K\"all\'en--Lehmann \textbf{spectral representation} 
under assumptions of the general principles: 
(i) the spectral condition, (ii) the Poincar\'e invariance and (iii) the completeness of the state space
\begin{align}
 \mathscr{D}(k^2) = \int_0 ^\infty d \sigma^2 \frac{\rho(\sigma^2)}{\sigma^2 - k^2} , \ k^2 \ge 0 , 
\label{eq:KL_spectral_repr-0} 
\end{align} 
with the weight function $\rho(\sigma^2)$ called the \textbf{spectral function} 
\begin{align}
   \theta(k_0) \rho(k^2) 
:=  (2\pi)^{d} \sum_{n }  |\langle 0 | \phi(0) | P_n \rangle|^2 \delta^D(P_n-k) .
\end{align}
The spectral function $\rho$ has contributions from a \textbf{stable single-particle state} with physical mass $m_{P}$ (pole mass) and intermediate many-particle states $| p_1,...,p_n \rangle$ with a continuous spectrum, 
\begin{align}
  \rho(k^2) =& Z \delta(k^2 -m_{P}^2) + \tilde\rho(k^2) , \ k^2 \ge 0 ,
\nonumber\\
\tilde\rho(k^2) =& (2\pi)^{d} \sum_{n=2}^{\infty}  |\langle 0 | \phi(0) | p_1,...,p_n \rangle|^2 \delta^D(p_1+...+p_n-k)
 .
 \label{spectral-f-R}
\end{align}
Then it is written as the sum of contributions from the {real pole} $k^2=m_P^2$ and the others $k^2 \in  [\sigma^2_{\rm min}, \infty)$
\begin{align}
 \mathscr{D}(k^2) =  \frac{Z}{m_P^2 - k^2} + \int_{\sigma^2_{\rm min}} ^\infty d \sigma^2 \frac{\tilde\rho(\sigma^2)}{\sigma^2 - k^2} 
 , \ k^2 \ge 0 . 
\label{eq:KL_spectral_repr-02} 
\end{align} 
 In the \textit{absence of complex  poles}, the spectral representation can be extended to the complex momentum $k^2 \in \mathbb{C}$. 
A propagator $\mathscr{D}(k^2)$ as a complex function of $z = k^2 \in \mathbb{C}$ has 
\begin{align}
 \mathscr{D}(k^2) &= \int_0 ^\infty d \sigma^2 \frac{\rho(\sigma^2)}{\sigma^2 - k^2}, \ k^2 \in \mathbb{C}-[\sigma^2_{\rm min}, \infty) , \quad
 \rho(\sigma^2)  := \frac{1}{\pi} \operatorname{Im} \mathscr{D}(\sigma^2+i\epsilon).
 \label{spectral-f-C}
\end{align} 
%
In the \textit{presence of complex poles}, 
the propagator has the \textbf{generalized spectral representation}
\begin{align}
& \mathscr{D}(k^2) 
 =  \mathscr{D}_{p}(k^2) + \mathscr{D}_{c}(k^2) , \ k^2 \in \mathbb{C}-([\sigma^2_{\rm min}, \infty) \cup \{ z_\ell \}_\ell ),
\nonumber\\
& \mathscr{D}_p(k^2)   :=  \frac{Z}{(v+iw) - k^2} + \frac{Z^*}{(v-iw) - k^2} , \quad Z  := \oint_{\gamma } \frac{d k^2}{2 \pi i} \mathscr{D}(k^2), \ 
\label{eq:spec_repr_complex} 
\nonumber\\ 
& \mathscr{D}_{c}(k^2) := \int_0 ^\infty d \sigma^2 \frac{\rho(\sigma^2)}{\sigma^2 - k^2} , \
 \rho(\sigma^2)  := \frac{1}{\pi} \operatorname{Im} \mathscr{D}(\sigma^2+i\epsilon) . 
\end{align}

\begin{figure}[t]
\centering
\includegraphics[width=7.5cm]{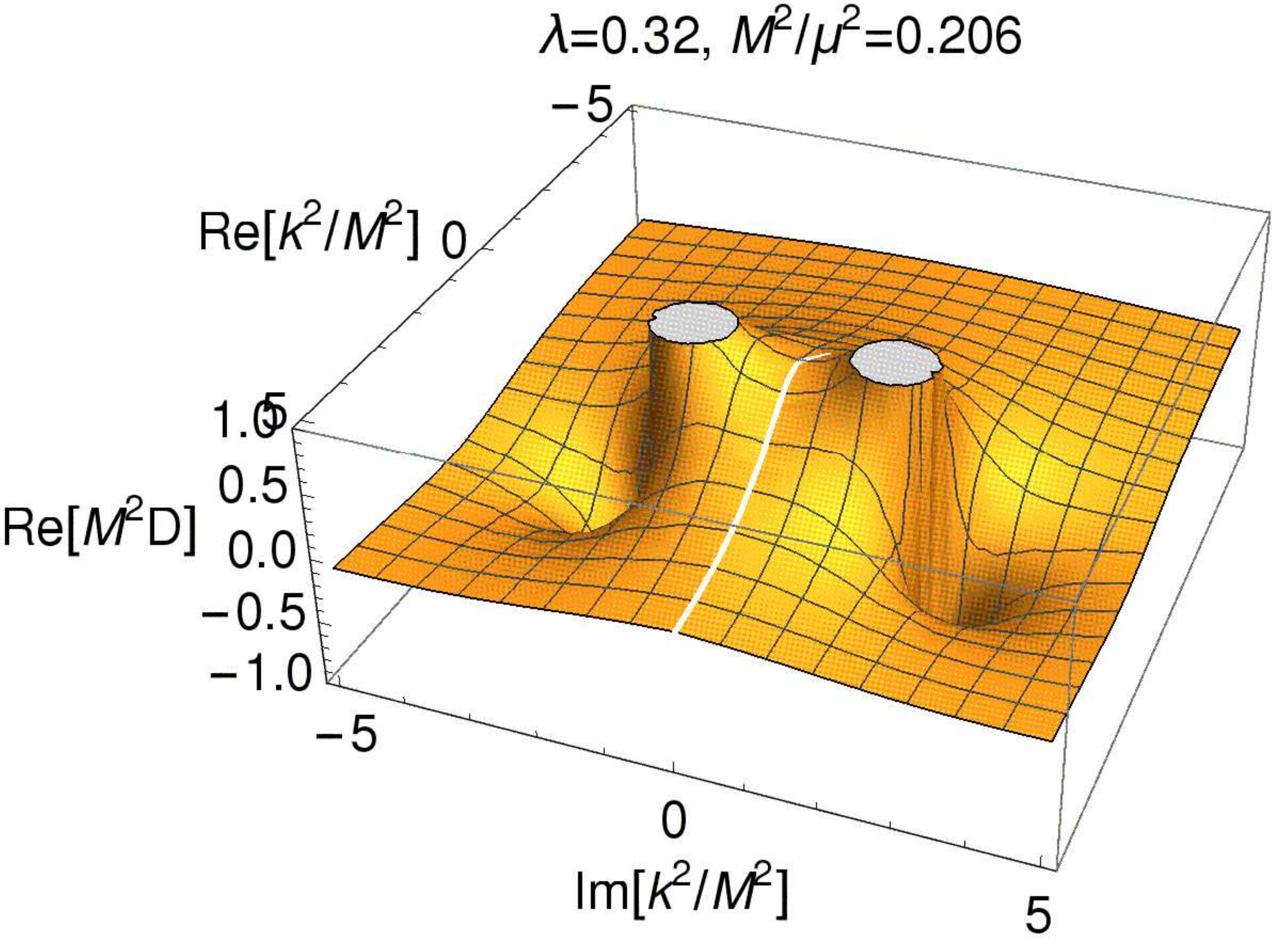}
\includegraphics[width=7.5cm]{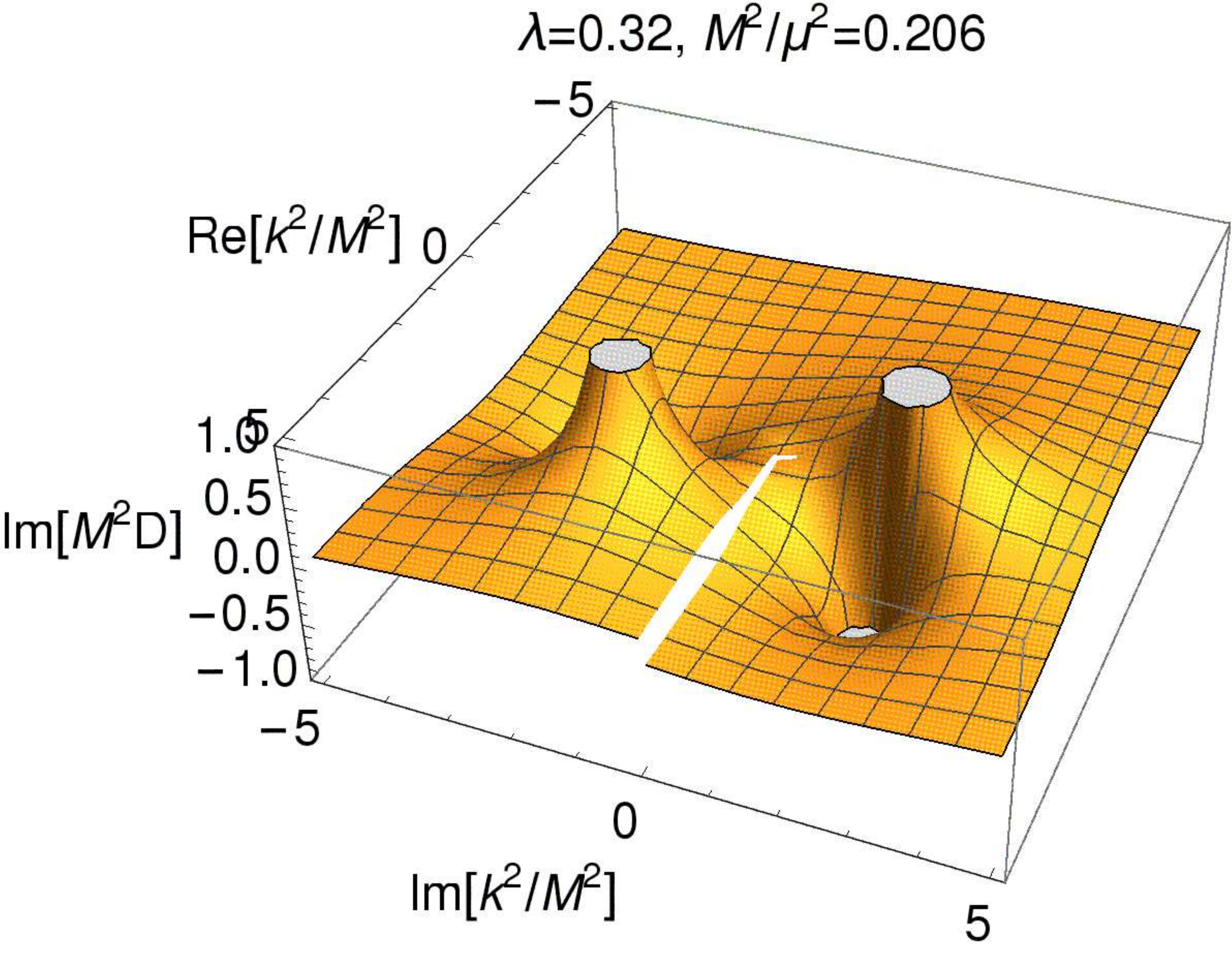}
\caption{
The gluon propagator $\mathscr{D}(k^2)$ as a complex function of  the complex squared momentum $k^2 \in \mathbb{C}$,  
(left) the real part $\operatorname{Re}\mathscr{D}(k^2)$, (right) the imaginary part $\operatorname{Im}\mathscr{D}(k^2)$, 
at the physical point of the parameters.  
}
\label{figPropAnaConPhys}
\end{figure}
\begin{figure}[t]
\centering
\includegraphics[width=7.5cm]{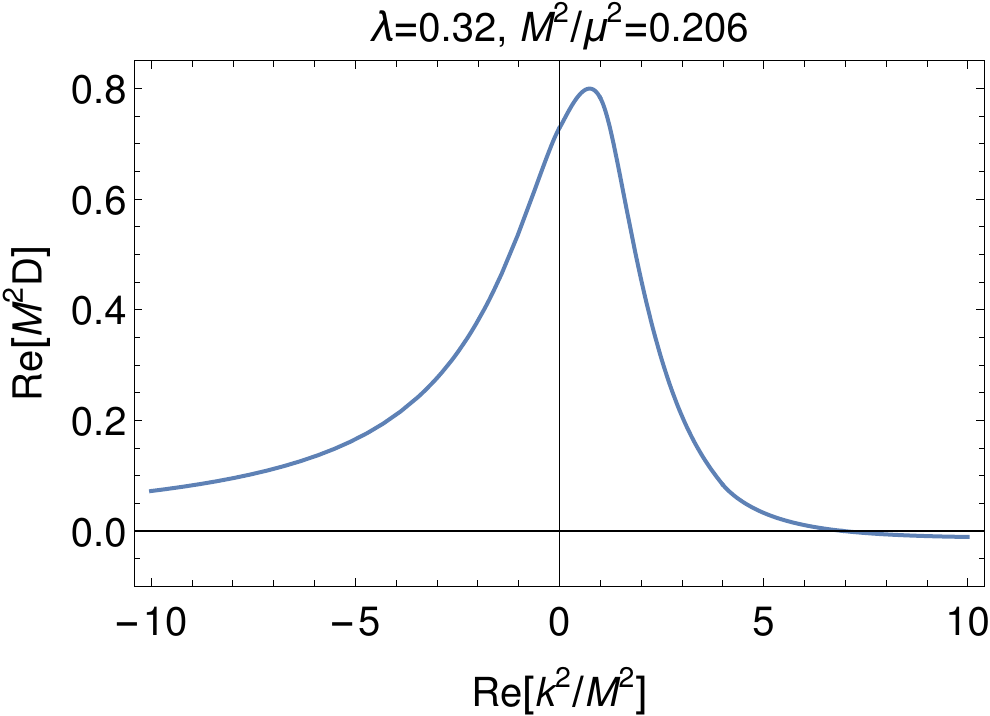}
\includegraphics[width=7.5cm]{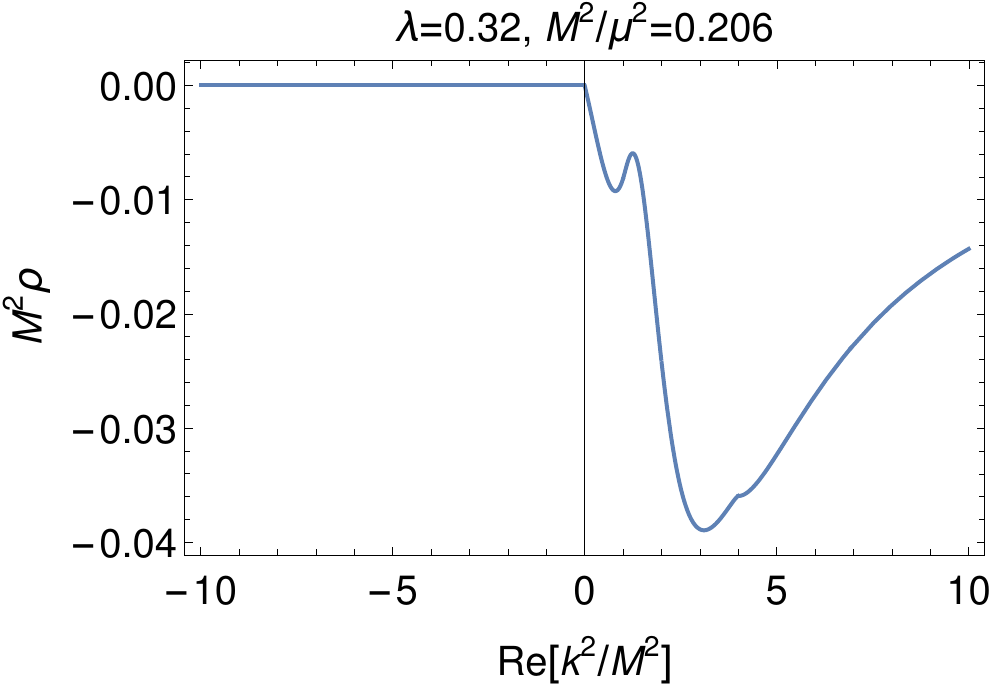}
\caption{
The gluon propagator $\mathscr{D}(k^2)$ as a function of $k^2 $ restricted on the real axis $k^2 \in \mathbb{R}$, 
(left) the real part $\operatorname{Re}\mathscr{D}(k^2)$, (right) the scaled imaginary part $\operatorname{Im}\mathscr{D}(k^2+i\epsilon)/\pi$ which is equal to the spectral function $\rho(k^2)$, 
at the physical point of the parameters .
}
\label{figPropRealAxisPhys}
\end{figure}

\textit{The gluon propagator $\mathscr{D}(k^2)$ in the massive Yang-Mills model has a pair of complex conjugate poles on the complex momentum on the complex momentum $k^2$ plane.} 
See Fig.~\ref{figPropAnaConPhys}. 
\textit{The spectral function $\rho(k^2)$ of the massive Yang-Mills model is always negative.} 
See Fig.~\ref{figPropRealAxisPhys}. 
The negativity of the spectral function and the existence of complex conjugate poles are interrelated \cite{HK19} :  
The negative spectral function  yields  \textbf{one pair of complex conjugate poles (or Euclidean real poles of multiplicity 2)}. 
This is consistent with negativity of $\rho(k^2)$ at large $k^2$ shown by Oehme and Zimmermann \cite{OZ80}.


\begin{figure}[t]
\centering
\includegraphics[width=7cm]{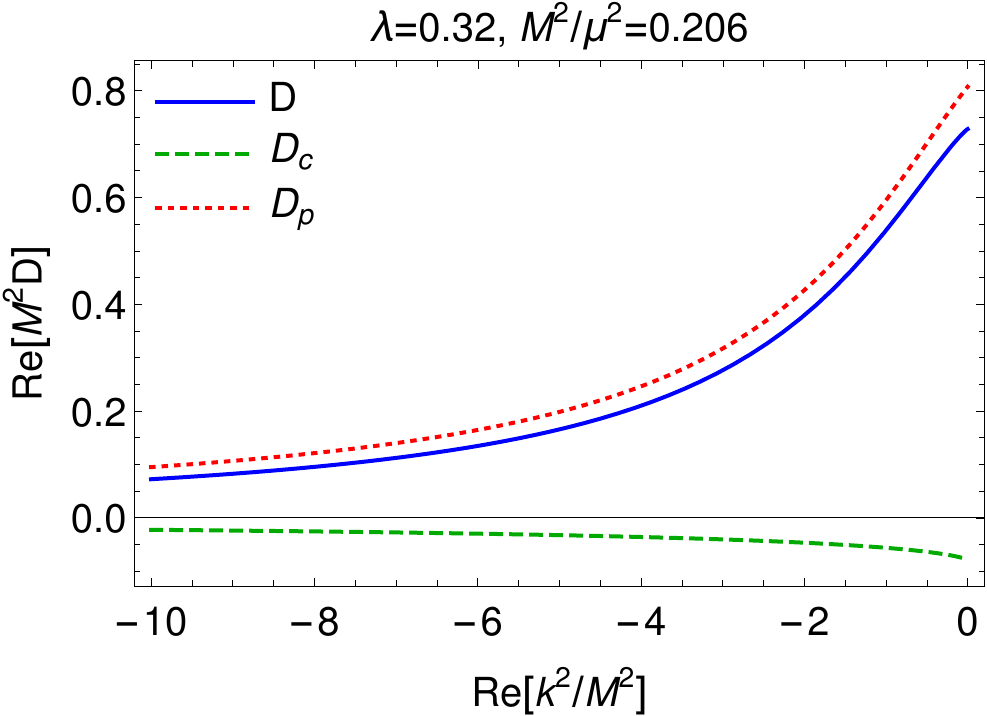}
\quad\quad
\includegraphics[width=7cm]{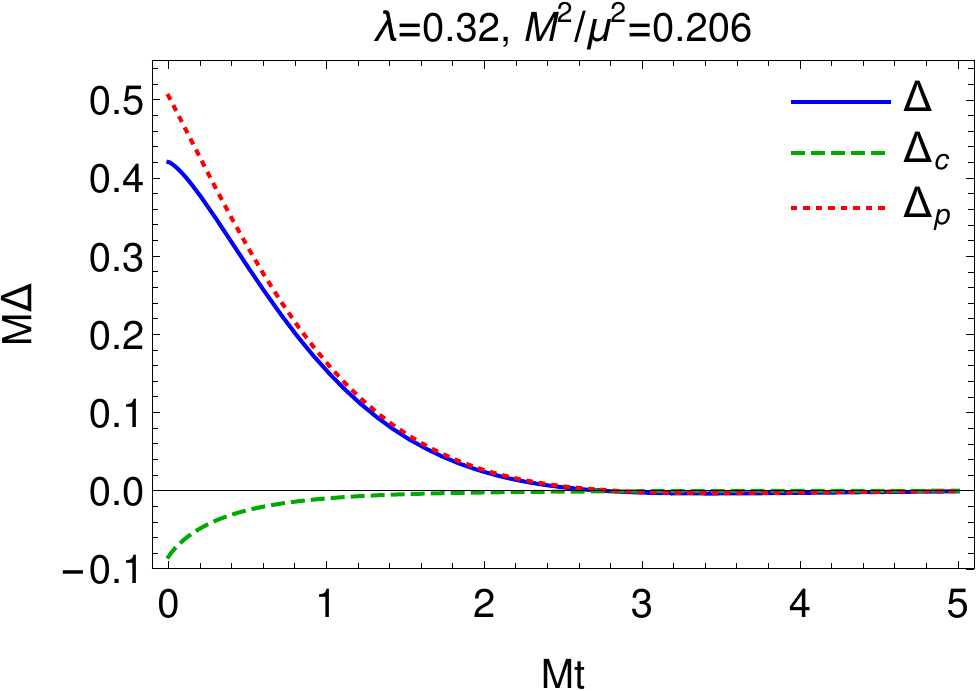}
\caption{\small
(left) the gluon propagator in the Euclidean region $\mathscr{D}(k_E^2)=\mathscr{D}_{p}(-k_E^2)+\mathscr{D}_{c}(-k_E^2)$ where $\mathscr{D}_{c}(-k_E^2)<0$
and (right) the gluon Schwinger function
  $\Delta(t)=\Delta_p(t)+\Delta_c(t)$ for $k^2=-k_E^2<0$.  
}
\label{figPropBreakPhys}
\end{figure}

The Schwinger function is also separated into the two parts (See Fig.\ref{figPropBreakPhys}.):
\begin{align}
\Delta(t)  =&  \Delta_{p}(t)  +  \Delta_{c}(t) , \quad
\Delta_{p,c}(t) := \int_{-\infty}^{+\infty} \frac{dk_E}{2\pi}  e^{ik_E t}  {\mathscr{D}}_{p,c}(-k_E^2) . \quad
\label{CCF-spec-Sch2m}
\end{align}
The cut part  $\Delta_{c}(t)$ is directly written as an integral of the spectral function as  
\begin{align}
 \Delta_{c}(t)  
=& \int_{-\infty}^{+\infty} \frac{dk_E}{2\pi}  e^{ik_E t}  \int_0 ^\infty d \sigma^2 \frac{\rho(\sigma^2)}{\sigma^2 + k_E^2}
=  \int_{0}^{\infty} d\sigma^2   \rho(\sigma^2) \frac{1}{2\sqrt{\sigma^2}} e^{-\sqrt{\sigma^2} t}  .
\label{Schwinger-f-cut}
\end{align}
To one-loop order in the massive Yang-Mills model, 
 the spectral function takes the negative value 
\begin{align}
  & \rho(\sigma^2)<0 \ \text{for} \ {}^{\forall} \  \sigma^2 > 0 
  \ \Rightarrow \ 
\Delta_{c}(t) < 0 \ \text{for} \  {}^{\forall} t \ge 0 ,
\label{negative-spec-f} 
\end{align} 
The pole part  $\mathscr{D}_p$ of the propagator  in the presence of a pair of complex conjugate poles at   
 $k^2 = v \pm i w (v,w>0)$ 
 with the  respective residues $Z, Z^* \in \mathbb{C}$ 
  reads in the Euclidean region  
\begin{align}
\mathscr{D}_p(-k_E^2) 
&= \frac{Z}{k_E^2 + (v + i w)} + \frac{Z^*}{k_E^2 + (v - i w)} 
 = 2 \frac{\operatorname{Re}[Z] k_E^2 + (v \operatorname{Re}[Z] + w \operatorname{Im}[Z])}{k_E^4 + 2 v k_E^2 + (v^2 + w^2)} .
\label{pp}
\end{align}
This pole part of the propagator agrees with the \textbf{Gribov-Stingl form} \cite{Stingl}. 
This is in good agreement with the lattice results. 
The pole part $\Delta_p(t)$ of the Schwinger function 
is exactly obtained as \cite{KWHMS19} 
\begin{align}
\Delta_p(t)
 =& \frac{\sqrt{\operatorname{Re}(Z)^2+\operatorname{Im}(Z)^2}}{ (v^2+w^2)^{1/4}  } \exp [ - t(v^2+w^2)^{1/4} \cos \varphi]  
 \cos  [ t(v^2+w^2)^{1/4} \sin \varphi + \varphi - \delta ]
. \
\label{exSchPole-s}
\end{align}
Therefore, the pole part has negative value at a certain value of $t$,
\begin{align}
\Delta_p(t) < 0 \ \text{for} \ {}^{\exists} \ t \ge 0 .
\end{align}
Thus, 
$\Delta (t) = \Delta_{c} (t)  + \Delta_{p} (t)$
has necessarily negative value at a certain value of $t$,
\begin{align}
\Delta(t) = \Delta_c(t) + \Delta_p(t) < 0 \ \text{for} \ {}^{\exists} \ t \ge 0 .
\end{align}
Thus we complete the proof that \textit{the reflection positivity is always violated in the massive Yang-Mills model to one-loop order}  (irrespective of the choice of the parameters $g$ and $M$).

\section{Conclusion and discussion}

In this talk we have investigated  the mass-deformed Yang-Mills theory in the covariant gauge obtained by just adding a gluon mass term to the Yang-Mills theory with the  Lorenz gauge fixing term and the associated   ghost term.
First, we have reconfirmed that the decoupling solution in the Landau gauge Yang-Mills theory is well reproduced from the mass-deformed Yang-Mills theory by taking into account loop corrections. 
 Second, we have shown violation of the reflection positivity as a necessary condition for gluon confinement for any value of the parameters in the mass-deformed Yang-Mills theory to one-loop quantum corrections, which follows from the existence of a pair of complex conjugate poles and the negativity of the spectral function for the gluon propagator. 

Moreover, it is shown that the mass-deformed Yang-Mills theory is obtained as a gauge-fixed version of the gauge-invariantly extended theory which is identified with the gauge-scalar model with a single fixed-modulus scalar field in the fundamental representation of the gauge group.
This equivalence is a consequence of the gauge-independent Brout-Englert-Higgs mechanism proposed recently by one of the authors \cite{Kondo16}. 
Thus we can discuss the implications for the existence of positivity violation/restoration crossover  in light of the Fradkin-Shenker continuity between Confinement-like and Higgs-like regions in a single confinement phase in the gauge-scalar model on the lattice.

\section*{Acknowledgement}

This work was supported by Grant-in-Aid for Scientific Research, JSPS KAKENHI Grant Number (C) No.15K05042 and No.19K03840.

\end{document}